\newcommand{\bld}[1]{\mbox{\boldmath{$#1$}}}
\newcommand{\ceqar}[2]{\parbox{4cm}{\begin{eqnarray*} #2 \end{eqnarray*}}
        \hfill\hspace{1cm}\parbox{1cm}
	{\begin{eqnarray}& &\nonumber\\ \label{#1}& & \end{eqnarray}\\}}

\documentstyle[preprint,aps,12pt,floats]{revtex}
\begin{document}
\draft
\tighten
\renewcommand{\floatpagefraction}{0.7}
\preprint{\hfill{\parbox{4.5cm}{TTP96-12\ \\ hep-ph/9605307\ \\
{\hfill April 1996}}}}

\title{BINARY BOOSTS}

\author{RUTH H\"ACKL\,, VIKTOR HUND AND HARTMUT PILKUHN}

\address{Institut f\"ur Theoretische Teilchenphysik, Universit\"at, Postfach
6980, D--76228 Karlsruhe, Germany}

\maketitle

\begin{abstract}{
The relativistic motion of an isolated two--body system  (bound or
unbound) of given lab energy $K^{0}$  in QED is 
separated into {\sl cms} motion and
relative motion. The relative motion equation ${\cal K}_{{\mbox{\tiny L}}} 
\psi_{{\mbox{\tiny L}}}
({\bf r}_{{\mbox{\tiny L}}} ) =0$ 
contains the momentum eigenvalue ${\bf K}$ of the
{\sl cms} motion. It is greatly simplified by a binary boost to the
atomic rest frame, where $K^{0}$ and ${\bf K}$ appear only in a 
Lorentz--invariant combination. This boost is not a product of 
single--particle boosts, which are useful only for perturbative 
interactions. CPT--invariance is demonstrated, and orthogonality
relations are derived.
}\end{abstract}

\section{Introduction}\label{einl}

This paper is about the relativistic two--body problem in QED. It treats 
the separation of the {\sl cms} motion 
from the relative motion in analogy with 
the nonrelativistic separation. The case of two spinless particles has been
treated previously \cite{mapi} but has no practical applications.
When one of the two particles is a lepton ($\mbox{e}$, $\mu$
or $\tau$) and the other is spinless, the two--body wave function
contains an additional boost:
\begin{equation}\label{gl1}
{\mbox A} = \left( \gamma + \gamma_{5} 
{\bf K}\bld{\sigma}_{1} / E \right) ^{1/2},
\qquad   \gamma = K^{0} / E  , \qquad {K^{0}}^{2} - {\bf K}^2 = E^{2}
\end{equation}
in units $\hbar\, =\, c\, =\,1 $. 
Here $K^{\mu}=\left( K^{0},{\bf K} \right)$ are
the eigenvalues of $P^{\mu}= \left(  i\partial_{t}, {\bf P} \right)$
with ${\bf P} = {\bf p}_{1} + {\bf p}_{2}\,$; $\bld{\sigma}_{1}$ are 
the Pauli matrices of particle 1, which are normally combined with the
Dirac matrix $\gamma_{5}$ into $\bld{\alpha}_{1} = \gamma_{5} 
\bld{\sigma}_{1}$. The {\sl binary boost} A resembles the boost 
${\mbox A}_{1}$ for a
single free particle:
\begin{equation}\label{gl2}
{\mbox A}_1 = \left( \gamma_{1} + \gamma_{5} {\bf K}_{1} \bld{\sigma}_{1}
/ m_{1} \right) ^{1/2}  , \qquad \gamma_{1} = K^{0}_{1} /m_{1} 
, \qquad {K^{0}_{1}}^{2} -{\bf K}_{1}^{2} = m_{1}^{2} .
\end{equation}
(An alternative form of ${\mbox A}_{1}$ which avoids $\bld{\sigma}_{1}$
under the square root is $(2 m_{1})^{-1/2} \left( K^{0}_{1} + m_{1} \right) ^{-1/2}$
$\left( K^{0}_{1} + m_{1} +\bld{\alpha}_{1}{\bf K}_{1} \right)$, but
(\ref{gl2}) is in fact more elegant.) These points will  be elaborated in 
section \ref{kleindirac}. They are of interest for muonic helium ($\mu^{-}\alpha$),
mu--pionium ($\mu^{-}\pi^{+}$) and also for precise recoil
corrections in ordinary  atoms.

Binary boosts are simplified by taking the $z$--axis along the total
momentum ${\bf K}\,$, $\bld{\sigma}_{1}{\bf K} = \sigma_{1z} K\,$, and
by abbreviating $K/E=\gamma v/c$ as ${\widehat K}$: $A = 
\left( \gamma + \gamma_{5}
{\widehat K} \sigma_{1z} \right) ^{1/2}$. 
The symbol $\beta$ will denote the {\sl parity Dirac matrix}
which will be taken in diagonal form:
\begin{equation}\label{t13}
\beta = \gamma_{0}  = \left( \begin{array}{cc}
1 & 0\\
0 & -1 \end{array} \right) \, , \qquad
\gamma_{5} = \left( \begin{array}{cc}
0 & 1 \\
1 & 0 \end{array} \right) \, . 
\end{equation}
The boost which we find for two spinor particles contains $\sigma_{z} =
\sigma_{1z} + \sigma_{2z}$ as well as a $\beta$--dependent mass ratio
$\mu_{\beta}$:
\begin{equation}\label{t14}
B= \left( 1 + \frac{1}{2} {\widehat K}^{2} 
\sigma_{z}^{2} + \mu_{\beta} {\widehat K}
\sigma_{z} \gamma \gamma_{5} \right) ^{1/2}\, , \quad {\widehat K} = K/E \, ,
\end{equation}
\begin{equation}\label{t15}
\mu_{\beta} = M_{-}/M_{+} \, , \quad M_{\pm} = m_{2} \pm \beta m_{1} \, .
\end{equation}
As $\sigma_{z}^{2}$ has only the two  eigenvalues
$4$ and $0$, an alternative form of $B$ is
\begin{equation}\label{t16}
B = {\tilde \gamma} + \mu_{\beta} {\widehat K} \sigma_{z} \gamma_{5} / 2 \, , 
\quad {\tilde \gamma} = \sqrt{1 + {\widehat K}^{2} \sigma_{z}^{2} / 4 }\, .
\end{equation}
Evidently, ${\tilde \gamma} = \gamma$ for $\sigma_{z}^{2} = 4$ and
${\tilde \gamma} = 1$ for $ \sigma_{z}^{2} = 0$. $B$ contains
two Pauli matrices but only one $\beta$ and one $\gamma_{5}\,$, i.e.~$B$ is 
an $8 \times 8$ matrix. For comparison, the two--body boost
$B_{12}$ used so far in perturbative QED is a $16 \times 16$--matrix,
\begin{equation}\label{t17}
B_{12} = A_{1} A_{2} = \left( \gamma_{1} + {\bf p}_{1}\bld{\alpha}_{1} /m_{1}
\right) ^{1/2} \left( \gamma_{2} + {\bf p}_{2}\bld{\alpha}_{2} / m_{2}
\right) ^{1/2} \, .
\end{equation}
It requires two separate conserved energies, $ K_{1}^{0}$ and $K_{2}^{0}$,
momentum operators ${\bf p}_{1}$ and ${\bf p}_{2}$ and Dirac matrices
$\beta_{1}, \beta_{2}, \gamma_{51} , \gamma_{52}$. The main advantage  
of $B$ is that it contains only the total momentum ${\bf K}$ 
instead of the individual momenta. It Lorentz transforms the asymptotic
part of the {\sl cms} spinorial wave function, the radial part of
which  is $e^{ikr}$ in spherical coordinates
(bound states have imaginary $k=i\kappa\,$; in QED, the
ground state has $\kappa_{0} = a_{B}^{-1}\,$, $a_{B}$ being the relativistic
Bohr radius). The boost $B_{12}$, on the other hand, requires 
asymptotic wave functions $e^{i{\bf k}_{1}{\bf r}_{1}}
e^{i{\bf k}_{2}{\bf r}_{2}}$ and then by analogy also two different
times, $e^{-iK_{1}^{0}t_{1}}e^{-iK_{2}^{0}t_{2}}$. 
The derivation of $B$ is given in the next section. 
As in the nonrelativistic case,
the shape of the potential 
$V({\bf r}_{{\mbox{\tiny L}}})$ is irrelevant in this 
derivation (${\bf r}_{{\mbox{\tiny L}}} = {\bf r}_{1} - {\bf r}_{2}$).
Anomalous magnetic moments can be included in the {\sl cms}
equation \cite{pi} but are as yet excluded 
from the boost.
Thus the present treatment 
applies to the bound
states of two charged leptons, which may be called leptonium (muonium
${\mbox e}^{-}\mu^{+}$, antimuonium ${\mbox e}^{+}\mu^{-}$,
positronium ${\mbox e}^{-}{\mbox e}^{+}$), as well as to
their scattering states up to $K^{0} \rightarrow \infty$ and\,/\,or $E
\rightarrow \infty$.

The relativistic separation of the {\sl cms} coordinate ${\bf R}$ reads 
\cite{mapi}
\begin{equation}\label{gl6}
{\bf r}_{1} = {\bf R} + \frac{1}{2} \left( 1 - \Delta m^{2}/E^{2} 
\right) {\bf r}_{{\mbox{\tiny L}}} , \qquad {\bf r}_{2} = {\bf R} - \frac{1}{2}
\left( 1 + \Delta m^{2}/E^{2} \right) {\bf r}_{{\mbox{\tiny L}}}
\end{equation}
with $\Delta m^{2} = m_{1}^{2} - m_{2}^{2} = -M_{+}M_{-}$. For loosely bound
states, $E^{2} \approx (m_{1} + m_{2})^{2}$ reduces 
(\ref{gl6}) to the nonrelativistic 
transformation except for the index {\small L} in ${\bf r}_{{\mbox{\tiny L}}}$ which still
signifies the lab system. This additional index is necessary because
${\bf r}_{{\mbox{\tiny L}}}$ is  
Lorentz--contracted in the direction ${\bf K}$.
The uncontracted {\sl cms} variable will be denoted by ${\bf r}^{\ast}$. In the
following, the $z$--axis is taken along ${\bf K}$, such that
$x_{{\mbox{\tiny L}}}$ and $y_{{\mbox{\tiny L}}}$ are 
not Lorentz--contracted, and
\begin{equation}\label{gl7}
z_{{\mbox{\tiny L}}} = z^{\ast}/\gamma , \qquad p_{{\mbox{\tiny L}}z} = \gamma p_{z}^{\ast} .
\end{equation}
The usual {\sl cms} potential $V^{\ast} = - \alpha / r^{\ast} $ ($\alpha = e^2$)
which is abstracted  from the QED Born approximation is 
Lorentz--contracted in the lab system. In the 16--component Dirac--Breit
equation which is applied successfully in atomic theory,
this contraction is not evident, but a careful analysis shows that it is
represented by the Breit operators \cite{mapi}. With a simple $V(r) = -\alpha /r$,
the {\sl cms} equation is free of Breit operators. However, this form
in fact requires  a third variable transformation which affects only
the distance, $ r \rightarrow r^{\ast}$. In other words, the Fourier transform
of the first Born approximation of the {\sl cms} scattering amplitude produces
a potential  $V(r)$ where $r$ is  a {\sl quasidistance} from the Dirac--Breit
point of view.
These procedures are up to now only approximately Lorentz--invariant and are
not repeated here.
We merely wish to reserve the index--free ${\bf r}$ for the 
quasiposition, which may or may not coincide with ${\bf r}^{\ast}$.

An important application of this 8--component formalism is to positronium. 
The relation $m_{2} = m_{1}$ simplifies
(\ref{gl6}) to ${\bf r}_{1} = {\bf R} + {\bf r}_{{\mbox{\tiny L}}}/2$, ${\bf r}_{2} = {\bf R}
-{\bf r}_{{\mbox{\tiny L}}}/2$. 
In this case, $B$ can be used in the small components ($ \beta = -1\, ,\,
\mu_{\beta} = \infty$) only for $ \sigma_{z}^{2} = 0\,$, in which case
the large components have $\sigma_{z}^{2} = 4\,$. The opposite case requires
a $\gamma_{5}$--transformation, which will be explained in section \ref{t2}.

We have found in the literature a formula which resembles (\ref{t14}):
Replacing $E \rightarrow m_{1}$ and $\mu_{\beta} \rightarrow 1\,$, (\ref{t14})
becomes identical with a boost for a single free particle of spin 1 \cite{wei}.

\section{Derivation of the 8--component equation}\label{t2}

A convenient starting point in the lab system is at present the 
16--component Dirac--Breit equation
\begin{equation}\label{t28}
\left( \pi^{0}_{{\mbox{\tiny L}}} - m_{1}\beta_{1} - m_{2}\beta_{2} - 
P_{{\mbox{\tiny B}}} \right) \psi_{{\mbox{\tiny DL}}} = 0 \, , \quad
\pi^{0}_{{\mbox{\tiny L}}} = i \partial_{t} - V_{12}
\end{equation}
where $P_{{\mbox{\tiny B}}}$ is the Dirac--Breit momentum operator \cite{mapi},
and the indices D and L stand for Dirac and lab system, respectively.
It contains ${\bf p}_{1}$ and ${\bf p}_{2}$ multiplied by Breit corrections
$1 - V_{12} \left( \bld{\alpha}_{1}\bld{\alpha}_{2} + \alpha_{1r}
\alpha_{2r} \right) / 2 $, which contain low--energy approximations.
In the future, one would like to formulate the Feynman rules for the
fermion--fermion scattering amplitudes in terms of 8--component
spinors, both in the lab system and in the {\sl cms}, and then derive
the interactions via Fourier transformations. At present,
this derivation is incomplete even in the {\sl cms}. The {\sl cms} equation
has been abstracted from (\ref{t28}), but most of its scattering
amplitudes have been checked against the Born approximation for
arbitrary energies. For the present construction of the free 
two--particle boost, we only need three properties of (\ref{t28}):
(i) it is translationally invariant in space and time,
\begin{equation}\label{t29}
i\partial_{t}\psi_{{\mbox{\tiny DL}}} = K^{0} \psi_{{\mbox{\tiny DL}}}\, ,
\quad\left( {\bf p}_{1} +{\bf p}_{2} \right) \psi_{{\mbox{\tiny DL}}} =
{\bf K} \psi_{{\mbox{\tiny DL}}} \, , 
\end{equation}
(ii) it is exact at all energies in
the asymptotic region $V_{12} = 0$ and (iii) it
contains only a single time $t$, which contributes a factor $e^{-i K^{0} t}$
for stationary states.
The resulting time--independent Dirac--Breit equation is again (\ref{t28}),
but with $\pi^{0}_{{\mbox{\tiny L}}} = K^{0} - V_{12}$. It contains the six
variables ${\bf r}_{1}$ and ${\bf r}_{2}$, precisely as in the
nonrelativistic case. For $V_{12}=0$, we thus have
\begin{equation}\label{t210}
\left( K^{0} - m_{1}\beta_{1} - m_{2} \beta_{2} - {\bf p}_{1}\bld{\alpha}_{1}
-{\bf p}_{2}\bld{\alpha}_{2} \right) \psi_{{\mbox{\tiny DL}}} = 0 \, .
\end{equation}
In the parity basis in which $\beta_{1}$ and $\beta_{2}$ are diagonal,
the components of $\psi_{{\mbox{\tiny DL}}}$ may be labelled by the index $g$ for a large 
component and $f$ for a small component of either particle: $\psi_{gg}$,
$\psi_{gf}$, $\psi_{fg}$ and $\psi_{ff}$.
In the following, these components are rearranged into sums
and differences  as follows:
\begin{equation}\label{t211}
\psi_{{\mbox{\tiny DL}}} = \left( \begin{array}{c} \psi_{g} \\
\psi_{f} \end{array} \right) = \frac{1}{\sqrt{2}}
\left( \begin{array}{c} \psi_{gg}+ \psi_{ff} \\
\psi_{gf} + \psi_{fg} \end{array} \right)
,\quad
\chi_{{\mbox{\tiny DL}}} = \left( \begin{array}{c} \chi_{g} \\
\chi_{f} \end{array} \right) = \frac{1}{\sqrt{2}} 
\left( \begin{array}{c} \psi_{gg} -\psi_{ff} \\
\psi_{gf} - \psi_{fg} \end{array} \right) .
\end{equation}
The index {\small D} stands for Dirac, the index {\small L} for the lab system.
The factor $2^{-1/2}$ makes the transformation unitary. The 16--component
equation thus assumes the form of two coupled 8--component equations,
\begin{equation}\label{t212}
\left( K^0 - \gamma_{5} p_{+} \right) \psi_{{\mbox{\tiny DL}}} 
= \beta M_{+} \, \chi_{{\mbox{\tiny DL}}}\, , \quad
\left( K^{0} + \gamma_{5} p_{-} \right) \chi_{{\mbox{\tiny DL}}}
 = \beta M_{+} \, \psi_{{\mbox{\tiny DL}}}\, ,
\end{equation}
\begin{equation}\label{t213}
p_{\pm} = {\bf p}_{1}\bld{\sigma}_{1} \pm {\bf p}_{2}\bld{\sigma}_{2}\, ,
\end{equation}
with $\beta , \gamma_{5}$ and $M_{+}$ defined in (\ref{t13}) and
(\ref{t15}). In words,
$\gamma_{5}$ exchanges the single index $g$ with the single index $f$ in $\psi$
and $\chi$, while $\beta$ multiplies $f$ by $-1$. The original
matrices $\beta_{1}$, $ \beta_{2}$, $\gamma_{51}$ and $\gamma_{52}$ are now
obsolete. Elimination of $\chi_{{\mbox{\tiny DL}}}$ by means 
of the first equation (\ref{t212})
yields
\begin{equation}\label{t214}
\left( K^{0} + \gamma_{5} p_{-} \right) \beta M_{+}^{-1} \left(
K^{0} - \gamma_{5} p_{+} \right) \psi_{{\mbox{\tiny DL}}} = \beta
M_{+} \psi_{{\mbox{\tiny DL}}} \, .
\end{equation}
Upon multiplication by $\beta M_{+}$, this becomes
\begin{equation}\label{t215}
\left( {K^{0}}^{2} - M_{+}^{2} - K^{0}p_{-}\mu_{\beta}^{-1} \gamma_{5}
- K^{0} p_{+} \gamma_{5} + \mu_{\beta}^{-1} p_{-}p_{+} \right)
\psi_{{\mbox{\tiny DL}}} = 0 . 
\end{equation}
The denominator $M_{-}= m_{2} - \beta m_{1}$ appears here because of
$\gamma_{5} \beta = - \beta \gamma_{5}$. At this point, we replace 
${\bf r}_{1}$ and ${\bf r}_{2}$ by ${\bf R}$ and ${\bf r}_{{\mbox{\tiny L}}}$
according to (\ref{gl6}), which leads to
\begin{equation}\label{t216}
{\bf p}_{1} = {\bf p}_{{\mbox{\tiny L}}} + \frac{1}{2} \left( 1 + \Delta
m^{2} / E^{2} \right) {\bf K}\,  ,\quad {\bf p}_{2}= -{\bf p}_{{\mbox{\tiny L}}}
+ \frac{1}{2} \left( 1 - \Delta m^{2} / E^{2} \right) {\bf K}
\end{equation}
where (\ref{t29}) has already been used, and $E^{2}= {K^{0}}^{2} - {\bf K}^{2}$
as in (\ref{gl1}). $\psi_{{\mbox{\tiny DL}}}$ is now a function of the
vector ${\bf r}_{{\mbox{\tiny L}}}$ and of the parameters $K^{0}$ and
$K$ (with the $z$--axis along ${\bf K}$). We now show that
$\psi_{{\mbox{\tiny DL}}}$ can be reduced to a function of the
components ${\bf r}_{{\mbox{\tiny L}}t} = ( x_{{\mbox{\tiny L}}}  ,\, 
y_{{\mbox{\tiny L}}})$ and $ z^{\ast} = \gamma z_{{\mbox{\tiny L}}}$
which contains $K^{0}$ and $K$ only in the combination ${K^{0}}^{2} - K^{2}
= E^{2}$. This justifies the choice of the transformation (\ref{gl6}),\,(\ref{gl7}) a posteriori.

The combinations $p_{\pm}$ of (\ref{t213}) become

\ceqar{t217}{\hspace{3cm}p_{-}& = &{\bf p}_{{\mbox{\tiny L}}}\bld{\sigma}_{t}+ \left( \gamma p_{z}^{\ast}
+ \frac{\Delta m^{2}}{2E^{2}} K \right) \sigma_{z} + \frac{1}{2} K
\Delta \sigma_{z} ,\\ \hspace{3cm}
p_{+}& =& {\bf p}_{{\mbox{\tiny L}}} \Delta \bld{\sigma}_{t} +
\left( \gamma p_{z}^{\ast} + \frac{\Delta m^{2}}{2E^{2}} K \right) \Delta
\sigma_{z} + \frac{1}{2} K \sigma_{z},}
\begin{equation}\label{t218}
\bld{\sigma} = \bld{\sigma}_{1} + \bld{\sigma}_{2}\, , \quad 
\Delta \bld{\sigma} = \bld{\sigma}_{1} - \bld{\sigma}_{2}\, , \quad
\bld{\sigma}_{t} = (\sigma_{x},
\sigma_{y} ) \, .
\end{equation}
Here we have denoted ${\bf p}_{{\mbox{\tiny L}}t}\bld{\sigma}_{t} =
{\bf p}_{{\mbox{\tiny L}}} \bld{\sigma}_{t}$ in order to save one index.
The operators ${\bf p}_{{\mbox{\tiny L}}t}$ and $p_{z}^{\ast}$ refer to the
{\sl cms}, 
but we avoid the notation ${\bf p}=-i\nabla$ which is reserved for the
quasimomentum.
In the Pauli spinor space $\chi_{1}\, \chi_{2}\,$, $\bld{\sigma}$ is
symmetric in 1 and 2. It transforms spin triplets into triplets and
annihilates spin singlets. $\Delta \bld{\sigma}$ is
antisymmetric and exchanges singlets 
$(\psi_{{\mbox{\tiny L}}s})$ with triplets
$(\psi_{{\mbox{\tiny L}}t})$. It turns out that
the following separation of mass factors is useful:
\begin{equation}\label{t219}
\psi_{{\mbox{\tiny DL}}} = \left( \begin{array}{c} \psi_{{\mbox{\tiny DL}}t} \\
\psi_{{\mbox{\tiny DL}}s} \end{array} \right) =
\left( \begin{array}{c} M_{+} \psi_{{\mbox{\tiny L}}t} \\
M_{-}\psi_{{\mbox{\tiny L}}s} \end{array} \right) .
\end{equation}
The index {\small D} disappears here; the new components $\psi_{{\mbox{\tiny L}}t}$
and $\psi_{{\mbox{\tiny L}}s}$ form an 8--component spinor $\psi_{{\mbox{\tiny
L}}}$, which satisfies the following equation:
\begin{eqnarray}\label{t220}
\lefteqn{
\bigg\lbrace {K^{0}}^{2} - M_{+}^{2} + \mu_{\beta}^{-1} p_{-}p_{+}
-K^{0} \bigg\lbrack 2 {\bf p}_{{\mbox{\tiny L}}} \bld{\sigma}_{1t} +
\left( 2 \gamma p_{z}^{\ast} + \frac{\Delta m^{2}K}{E^{2}} \right)
\sigma_{1z} +} \hspace{2cm}\nonumber\\
& &\hspace{3.2cm} +  \frac{K}{2} \left( \mu_{\beta}^{-1} \Delta 
\sigma_{z} + \mu_{\beta} \sigma_{z} \right)\bigg\rbrack  \gamma_{5}
\bigg\rbrace \psi_{{\mbox{\tiny L}}} = 0 .
\end{eqnarray}
Here we have used $\bld{\sigma} + \Delta \bld{\sigma} = 2   
\bld{\sigma}_{1}$. The expression $p_{-}p_{+}$ simplifies considerably,
\begin{equation}\label{t221}
p_{-}p_{+} = \frac{\Delta m^{2}}{E^{2}} K^{2} + 2 K \gamma p_{z}^{\ast} .
\end{equation}
Remembering $\Delta m^{2} = -M_{+}M_{-}$, the form $M_{+} \Delta 
m^{2} K^{2}/ E^{2} M_{-} = - M_{+}^{2} K^{2}/E^{2}$ is combined with
$-M_{+}^{2}$ into $-M_{+}^{2}(1+K^{2}/E^{2}) = - M_{+}^{2} \gamma ^{2} \,$,
 $\gamma = K^{0}/E$.
One may thus remove one factor $\gamma$ from (\ref{t220}). The 
result is ${\cal K}_{{\mbox{\tiny L}}} 
\psi_{{\mbox{\tiny L}}} = 0$, where
\begin{eqnarray}\label{t222}
\lefteqn{{\cal K}_{{\mbox{\tiny L}}} = (E^{2} -M_{+}^{2} ) \gamma + 
2 K \mu_{\beta}^{-1}p_{z}^{\ast}
- \Big\lbrack 2 E {\bf p}_{{\mbox{\tiny L}}} \bld{\sigma}_{1t} +
( 2 E \gamma p_{z}^{\ast} - M_{+}M_{-} \widehat K ) 
\sigma _{1z} +}\nonumber
\hspace{4cm}\\
& &\hspace{1.6cm} 
+\frac{1}{2} {\widehat K}E^{2} \left( \mu_{\beta}^{-1} \Delta\sigma_{z} +
\mu_{\beta} \sigma_{z} \right) \Big\rbrack \gamma_{5} \, ,
\end{eqnarray}
with ${\widehat K} = K/E$ as before.
In the {\sl cms}, i.e.~${\widehat K}=0\,$, $\gamma =1$, this reduces to 
\begin{equation}\label{t223}
{\cal K} \psi = 0\, , \quad {\cal K} = E^{2} - M_{+}^{2} - 2 E (
{\bf p}_{{\mbox{\tiny L}}} \bld{\sigma}_{1t} + p_{z}^{\ast} \sigma_{1z} )
\gamma_{5}\, .
\end{equation}
The operator ${\cal K}$ is independent of $\bld{\sigma}_{2}$ and
well--behaved
for $m_{1}=m_{2}$, where $M_{+}^{2} = 2 m_{1}^{2} ( 1 +\beta )$
annihilates the small components.
It remains to find a transformation $\psi_{{\mbox{\tiny L}}} = B \psi$
which reduces ${\cal K}_{{\mbox{\tiny L}}}$ to ${\cal K}$ after multiplication
by another matrix ${\bar B}$ from the left
\begin{equation}\label{t224}
{\bar B} {\cal K}_{{\mbox{\tiny L}}} B \psi = 
{\cal K} \psi =0 \, ,\quad {\cal K} = {\bar B} 
{\cal K}_{{\mbox{\tiny L}}} B \, .
\end{equation}
As the operator $ -2E{\bf p}_{{\mbox{\tiny L}}}\bld{\sigma}_{1t}$
occurs in (\ref{t223}) in the same form as in (\ref{t222}), one needs
\begin{equation}\label{t225}
{\bar B} \bld{\sigma}_{1t} = \bld{\sigma}_{1t} B^{-1} .
\end{equation}
The form (\ref{t222}) suggests that $B$ may not contain $\bld{\sigma}_{1t}$
and $\bld{\sigma}_{2t}$. Next, one may note that  
\begin{equation}\label{t226}
\sigma_{1z}\bld{\sigma}_{1t} = - \bld{\sigma}_{1t} \sigma_{1z}\, , \quad
\Delta \sigma_{z} \bld{\sigma}_{1t} = - \bld{\sigma}_{1t} \sigma_{z}.
\end{equation}
After a few fruitless attempts, one solves (\ref{t224}) with
\begin{eqnarray}\label{t227}
B&=&\left( 1 + \frac{1}{2} {\widehat K}^{2} \sigma_{z}^{2} + \mu_{\beta}
{\widehat K}
 \sigma_{z} \gamma \gamma_{5} \right) ^{1/2}\, ,\quad
B^{-1}= \left( 1 + \frac{1}{2} {\widehat K}^{2} \sigma_{z}^{2} -\mu_{\beta}
{\widehat
K}
\sigma_{z} \gamma\gamma_{5} \right) ^ {1/2},
\\
\label{t228}
{\bar B}& =& \left( 1 + \frac{1}{2} {\widehat K}^{2} \Delta 
\sigma_{z}^{2} + \mu_{\beta}^{-1}{\widehat K}
 \Delta \sigma_{z} \gamma \gamma_{5} \right) ^{1/2}\, .
\end{eqnarray}
It remains to show how the eigenvalue $\infty$ of $\mu_{\beta}$ (\ref{t15})
is avoided for positronium. When the large components have $\sigma_{z}^{2} =
4\,$, $\sigma_{z} = \pm 2$ implies $m_{l} = m_{j} \pm 1$
($m_{l}$ and $m_{j}$ are the eigenvalues of $L_{z}$ and $J_{z}=L_{z} +
\sigma_{z} / 2$).
These states have parity $(-1)^{j+1}\,$ ($J^{2} = j (j+1)$). The
orbital parity of the small components is opposite, i.e.~$(-1)^{j}$.
The spin function can be either singlet or triplet, and in either case
$m_{l} = m_{j}$ ensures $\sigma_{z} = 0\,$, i.e.~the combination $\mu_{\beta}
\sigma_{z}$ can be taken to vanish in (\ref{t16}).
Thus in the application to positronium, one must discuss the spin structure 
for  $m_{1} \not= m_{2}$ and then take $m_{1} = m_{2}$ in the
final forms. When the large components have $\sigma_{z} = 0$, a chiral
transformation $ \psi_{{\mbox{\tiny DL}}} = \gamma_{5} \psi_{{\mbox{\tiny
DL}},ch}$ transforms $\mu_{\beta}$ into $\mu_{\beta}^{-1}$, such that the
above argument applies again.

\section{The Klein--Dirac boost}\label{kleindirac}

The 4--component Dirac--Klein--Gordon--Breit equation for systems such
as $\mu^{-}\pi^{+}$ reads \cite{mapi}
\begin{equation}\label{t324}
\left( {\pi^{0}_{{\mbox{\tiny L}}}}^{2} +{\cal K}_{1} -{\cal K}_{2}
- 2 m_{1} \pi_{{\mbox{\tiny L}}} ^{0} 
\beta \right) \psi_{{\mbox{\tiny L}}}'  = \gamma_{5} \left( \lbrace {\bf p}_{1}
 , \pi_{{\mbox{\tiny L}}}^{0} \rbrace \bld{\sigma}_{1} + b_{1} \right) 
\psi_{{\mbox{\tiny L}}} ' 
\end{equation}
where $b_{1}$ is the Breit modification of the momentum operator,
\begin{equation}\label{t325}
b_{1} = - V_{12} \left( \bld{\sigma}_{1} {\bf p}_{2} + \sigma_{1r}
p_{2r} \right) \, ,
\end{equation}
and ${\cal K}_{i} = m_{i}^{2} + p_{i}^{2}$. The coordinate transformations
(\ref{gl6}) and (\ref{gl7}) give
\begin{equation}\label{t326}
{\cal K}_{1} - {\cal K}_{2} = \gamma^{2} \Delta m^{2} + 2 \gamma p_{z}^{\ast} K \, .
\end{equation}
Setting now
$V_{12}= 0$ and extracting one factor $\gamma\,$, one
obtains ${\cal K}_{{\mbox{\tiny L}}}' \psi_{{\mbox{\tiny L}}}' = 0\, $, with
\begin{equation}\label{t327}
{\cal K}_{{\mbox{\tiny L}}}' = \left( E^{2} + \Delta m^{2} \right)
\left( \gamma - \gamma_{5} {\widehat K}\sigma_{1z} \right) + 2 K p_{z}^{\ast} -
2 \gamma_{5} E \left( \gamma p_{z}^{\ast}\sigma_{1z} + {\bf p}_{{\mbox{\tiny
L}}} \bld{\sigma}_{1t} \right) - 2 m_{1} \beta E \, .
\end{equation}
Writing the lab spinor $\psi_{{\mbox{\tiny L}}}$ as a boost $A$ times the
{\sl cms} spinor $\psi '\,$,
 $\psi_{{\mbox{\tiny L}}}' = A \psi ' \,$, one finds
\begin{equation}\label{t328}
{\cal K}' \psi ' = 0 \, ,\quad {\cal K}' = A\, {\cal K}_{{\mbox{\tiny L}}}' \,
A \, , \quad A = \left( \gamma + \gamma_{5} {\widehat
 K} \sigma_{1z} 
\right) ^{1/2},
\end{equation}
\begin{equation}\label{t329}
{\cal K}' = E^{2} + \Delta m^{2} - 2 E m_{1} \beta - 2 E \left( p_{z}^{\ast}
\alpha_{z} + {\bf p}_{{\mbox{\tiny L}}} \bld{\alpha}_{t} \right) \, .
\end{equation}
The asymptotic energy $E_{1}$ of the spinor particle is
$E_{1} = (E^{2} + \Delta m^{2} ) / 2E\,$, such that the first two terms in 
(\ref{t329}) arise from $2E E_{1}$. As ${\cal K}'$ has the Dirac
operator structure, it is clear that its eigenvalues depend on
$ E^{2} + \Delta m^{2} $ and $ - 2 E m_{1} \beta$ only via
$(E^{2} + \Delta m^{2} ) ^{2} - 4 m_{1}^{2} E^{2}$.
Setting now
\begin{equation}\label{t330}
\psi ' = \left( 2 k^{2} E \right) ^{-1/2} \left( E_{1} - \beta
m_{1} \right) ^{1/2} \left( E^{2} - M_{-}^{2} \right) ^{1/2} \psi\, ,
\quad k^{2} = E_{1}^{2} - m_{1}^{2}
\end{equation}
one finds that ${\cal K }'$ is transformed into ${\cal K}$ as given in
(\ref{t223})
(the resulting boost is rather different, however).
The point $k^{2}=0$ comprises two thresholds at 
$E^{2} = (m_{1} + m_{2} )^{2}$ and two at 
$E^{2} = ( m_{1} - m_{2} )^{2} $. In our example of a $\mu^{-}\pi^{+}$
system, $E=m_{1} + m_{2}$ is the $\mu^{-}\pi^{+}$ threshold,
$E= -m_{1} - m_{2}$ the $\mu^{+}\pi^{-}$ threshold, $E=m_{2} -m_{1}$ the
$\mu^{+}\pi^{+}$ threshold and $E = m_{1} -m_{2}$ the $\mu^{-}\pi^{-}$
threshold. For $m_{1} = m_{2}$, one factor $E$ can be
separated from (\ref{t329}), leading to
${\cal K}' = E - 2m_{1} \beta -2 ( p_{z}^{\ast} \alpha_{z} +
{\bf p}_{{\mbox{\tiny L}}}
\bld{\alpha}_{t} )\,$. This completely eliminates the doubly--charged
channels, which now require a separate equation.
A similar decoupling occurs in the double--Dirac case for $m_{1} = m_{2}\,$,
which is discussed below.

\section{Orthogonality relations}\label{scalar}

A single 4--component Dirac spinor has components
$\psi_{{\mbox{\tiny R}}}$ and $\psi_{{\mbox{\tiny L}}}$ in the chiral
basis, where $\gamma_{5}$ is diagonal. $\beta$ and $\gamma_{5}$ simply
change their places in (\ref{t13}), i.e.~$\beta$ exchanges
$\psi_{{\mbox{\tiny R}}}$ and $\psi_{{\mbox{\tiny L}}}$. As $\beta$ is
part of the parity transformation, $\psi_{{\mbox{\tiny R}}}$ and
$\psi_{{\mbox{\tiny L}}}$ are not parity eigenstates. They do form
separate representations of the Lorentz group, however.

In the present case, both $\psi$ and $\chi$ can have $\beta$
diagonal and remain separate under Lorentz transformations. Elimination
of $\psi_{{\mbox{\tiny DL}}}$ from (\ref{t212}) leads to an equation for
$\chi_{{\mbox{\tiny DL}}}$ in which the two brackets of (\ref{t214}) 
are interchanged, which is equivalent to the substitution $p_{+} 
\longleftrightarrow - p_{-}$.
The equation
corresponding to (\ref{t215}) for $\chi_{{\mbox{\tiny DL}}}$ is thus
\begin{equation}\label{t338}
\left( {K^{0}}^{2} - M_{+}^{2} + \left( \mu_{\beta}^{-1} p_{+} + p_{-}
\right) K^{0} \gamma_{5} + \mu_{\beta}^{-1} p_{+} p_{-} \right) 
\chi_{{\mbox{\tiny DL}}} = 0\, .
\end{equation}
It will be shown in the following that the orthogonality relations
require both $\psi$ and $\chi$.
The mass separation analogous to (\ref{t219}) is
\begin{equation}\label{t339}
\chi_{{\mbox{\tiny DL}}}  = \beta
\left( \begin{array}{r}  \chi_{_{{\mbox{\tiny L}}t}} \\
\mu_{\beta}^{-1}\chi_{_{{\mbox{\tiny L}}s}} \end{array} \right) .
\end{equation}
The relation analogous to (\ref{t222}) is ${\cal K}_{\chi\!_{\mbox{\tiny{ L}}}}
\chi_{_{\mbox{\tiny L}}} = 0\,$, with
\begin{eqnarray}\label{t340}
\lefteqn{{\cal K}_{\chi_{\mbox{\tiny L}}} = (E^{2} -M_{+}^{2} ) \gamma + 
2 K \mu_{\beta}^{-1}p_{z}^{\ast}
- \Big\lbrack 2 E {\bf p}_{{\mbox{\tiny L}}} \bld{\sigma}_{1t} +
( 2 E \gamma p_{z}^{\ast} - M_{+}M_{-} \widehat K ) 
\sigma _{1z} +}\nonumber
\hspace{4cm}\\
& &\hspace{2.5cm} 
+\frac{1}{2} {\widehat K}E^{2} \left( \mu_{\beta} \Delta\sigma_{z} +
\mu_{\beta}^{-1} \sigma_{z} \right) \Big\rbrack \gamma_{5} \, ,
\end{eqnarray}
The boost is now different, $\chi_{_{\mbox{\tiny L}}} = B_{\chi} \chi$, where
\begin{equation}\label{t341}
B_{\chi} = \left( 1 +\frac{1}{2} {\widehat K}^{2} \Delta \sigma_{z}^{2}
+\mu_{\beta} {\widehat K} \Delta \sigma_{z} \gamma \gamma_{5} \right) ^{1/2}
= {\bar B}^{\dagger}
\end{equation}
where ${\bar B}$ has been given in (\ref{t228}). The hermitian conjugation
of ${\bar B}$ in (\ref{t228}) merely exchanges the positions of
$\mu_{\beta}^{-1}$ and $\gamma_{5}$. Observing $\left(\mu_{\beta}^{-1}
\gamma_{5} \right)^{\dagger}= \gamma_{5} \mu_{\beta}^{-1} =
\mu_{\beta}\gamma_{5}\,$, one readily verifies (\ref{t341}).
The coordinate transformation (\ref{gl6}) implies $d\,^{3}r_{1}\, 
d\,^{3}r_{2} =
d\,^{3}r_{{\mbox{\tiny L}}}\, d\,^{3}\!R\,$, and the orthogonality of states with
different momenta ${\bf K}$ follows simply from
$\int d^{3}\!R\,\, \mbox{exp} \lbrace - i ({\bf K} - {\bf K}') {\bf R} \rbrace =
(2\pi )^{3} \delta ({\bf K} -{\bf K}' ) $.
A subsequent boost allows one to discuss the orthogonality of the
functions $\psi ({\bf r}_{{\mbox{\tiny L}}} )$ in the {\sl cms}.
The relevant equations here are ${\cal K} \psi = 0$ and ${\cal K}\chi = 0\,$,
with ${\cal K}$ given by (\ref{t223}). The latter equation follows from
${\cal K}_{\chi_ {\mbox{\tiny L}}} \chi_{_{\mbox{\tiny L}}} = 0$ for
${\widehat K} =0\, $, $\gamma=1\,$, i.e.~${\cal K}_{\chi} = {\cal K}$
in the {\sl cms}.

In the ordinary Dirac equation, neither the potential $V(r)$ nor the
momentum ${\bf p} = - i \nabla$ appear in the orthogonality relations.
Both properties are also achieved here, but the details are surprising.
The general operator form of ${\cal K}$ will be needed for all 
values of $r$, not just for $r \rightarrow \infty$ where $V(r)=0$.
This form can be derived from the Fourier transform of the QED
Born approximation for elastic scattering $1 + 2 \rightarrow 1' + 2'$ in the
{\sl cms}:
\begin{equation}\label{orth42}
{\cal K} = E^{2} - M_{+}^{2} - 2 E \bld{\alpha} {\bf p} - 2 E V(r) - \Lambda
V'(r)\, ,
\end{equation}
\begin{equation}\label{orth43}
\bld{\alpha} = \gamma_{5} \bld{\sigma}_{1}\, , \quad
\Lambda = i \alpha_{r} \bld{\sigma}_{1}\bld{\sigma}_{2} = ( \bld{\alpha}
\times \bld{\sigma}_{2} ) _{r} + i \gamma_{5} \sigma_{2r}\, ,
\end{equation}
and $V' = dV / dr\,$, $ \sigma_{r} = \bld{\sigma}\hat{\bf{r}}$ 
as usual. The derivation from the Dirac--Breit
equation leads to a more complicated form in the variable ${\bf r}^{\ast}
= (z^{\ast}, x_{{\mbox{\tiny L}}}, y_{{\mbox{\tiny L}}} )\,$, which involves
Breit operators and also $V^{2}(r^{\ast})$. However, it turns out that the
substitution
\begin{equation}\label{orth44}
r^{\ast} \approx r + \alpha / 2 E
\end{equation}
does reduce the Dirac--Breit expression to (\ref{orth42}) at low energies. As
the QED Born approximation is free of low--energy approximations,
(\ref{orth42})
is much more practical than the Dirac--Breit equation.

To derive the orthogonality relations for the solution of a Hamiltonian
equation $H\psi = E\psi\,$, one writes $H \psi_{j} = E_{j} \psi_{j}\,$,
$(H\psi_{i} )^{\dagger} = E_{i} \psi_{i}^{\dagger} \,$, multiplies the first
equation by $\psi_{i}^{\dagger}\,$, the second one by $\psi_{j}\, $,
subtracts the second product from the first one, and integrates
over all configuration space:
$(E_{i} - E_{j})\int \psi_{i}^{\dagger}\psi_{j} =0$. The
method can also be applied to the Klein--Gordon equation, $\lbrack
(E - V(r)\,)^{2} + \nabla^{2} - m^{2} \rbrack \psi = 0 $ (in units $\hbar
=c = 1$), but the square of $E - V$ produces a weight $w_{ij}= E_{i}+E_{j}
-2V$,
i.e.~$\int \psi_{i}^{\ast}\, w_{ij}\, \psi_{j} = \delta_{ij}\,$.
The case at hand is more complicated because $\Lambda$ is not hermitian. Its
hermitian part is the recoil--corrected hyperfine operator. Its antihermitian
part has zero expectation values in fine structure eigenstates; it is needed in
positronium where fine and hyperfine structures are comparable (the energy
eigenvalues remain real). Fortunately, 
the equation ${\cal K}_{\chi} \, \chi = 0$ 
has $\Lambda$ replaced by $\Lambda ^{\dagger}\,$:
\begin{equation}\label{orth45}
{\cal K}_{\chi} = E^{2} - M_{+}^{2} - 2E \bld{\alpha}{\bf p} - 2 E V(r) -
\Lambda^{\dagger} V'(r)\, .
\end{equation}
One may thus envisage orthogonality relations $\int \chi _{i}^{\dagger}\, 
w_{ij}\, 
\psi_{j} = \delta_{ij}\,$, where the antihermitian component disappears
with $\Lambda^{\dagger\dagger} - \Lambda =0$.
However, the factor $E$ multiplies not only $V(r)$ as in the
Klein--Gordon equation, but also $\bld{\alpha} {\bf p}$. To avoid ${\bf p}$
in the orthogonality relations, $E$ must be divided off. But
then $\Lambda / E$ is both non--hermitian and energy--dependent, in 
which case $V'$ will remain in the orthogonality relations.

For $V= - \alpha / r$, one may introduce a dimensionless scaled
variable,
\begin{equation}\label{orth46}
{\widetilde r} = E r\, \quad \partial / \partial {\widetilde r} = E^{-1}
\partial / \partial r \, \quad {\widetilde {\bf p}} = {\bf p}/ E\, ,
\end{equation}
and divide ${\cal K}$ by $E^{2} = s$:
\begin{equation}\label{orth47}
\left( 2 \bld{\alpha} {\widetilde{\bf p}} + 2 V ( {\widetilde r}\,) + \Lambda
V' ( {\widetilde r}\,) - 1 + M_{+}^{2} / s 
\right) \psi ( {\widetilde r}\,) = 0 \, .
\end{equation}
Using the corresponding equation for $\chi^{\dagger}$, one arrives at
\begin{eqnarray}
(s_{i}^{-1} - s_{j}^{-1} ) \int \chi^{\dagger}_{i}\,  M_{+}^{2} \, \psi_{j}\,
d^{3}{\widetilde r} & = & 0 \, ,  \quad s_{i} = E_{i}^{2}\, , \label{orth48}\\
\int \chi^{\dagger}_{i} \, M_{+}^{2}\,
 \psi_{j}\, d^{3}{\widetilde r} & = &\delta_{ij}
\, . \label{orth49}
\end{eqnarray}
Remembering $M_{+}^{2} = m_{1}^{2} + m_{2}^{2} + 2 m_{1}m_{2} \beta\,$, this
is a simple generalization of the static limit $m_{1}/m_{2} = 0$.
For positronium, the small components do not contribute to (\ref{orth49}).

Equation (\ref{orth48}) is explicitly CPT--invariant: Every
bound state $s_{i}$ has two different eigenvalues $E_{i}\,$, namely
$E_{i} = \sqrt{s_{i}} \equiv m_{{\mbox{\tiny A}}i}$ and $ E_{i} = 
- \sqrt{s_{i}} \equiv - m_{{\mbox{\tiny A}}i}\,$, where $m_{{\mbox{\tiny A}}i}$
denotes the atomic mass in the state $i$ (an excited atom is heavier
than its ground state).
Returning now to the time--dependent {\sl cms} equation with $i \partial _{t}
=E\,$, one finds that $E_{i} = - m_{{\mbox{\tiny A}}i}$ belongs to the
antiatom of mass $m_{{\bar{\mbox{\tiny A}}}i}\,$, i.e.~$m_{{\bar{\mbox{\tiny A}}}i}
= m_{{\mbox{\tiny A}}i}\,$. Positronium is its own
antiatom, of course.

Vacuum polarization introduces an extra scale into $V(r)$.
The equation remains CPT--invariant, but (\ref{orth48}) becomes more
complicated. In the Klein--Dirac case of section \ref{kleindirac}, the
spinless particle normally has an extended charge distribution, leading to
$V(r) \not= - \alpha/r$ for small $r$. The simplest way out
is of course to take $V({\widetilde r}\,)$ as an arbitrary function
of ${\widetilde r}$ in (\ref{orth47}). In any case, finite charge
distribution models need some tuning in order to conform with CPT.

It should be stated that the range of the dimensionless radial variable
${\widetilde r}$ is $0 < {\widetilde r} < \infty$ both for
atoms and for antiatoms. In the old variable $r$, antiatoms have
{\sl negative distances}. Of course, this makes as little sense
as the claim that antiatoms {\sl fly backwards in time}.

We conclude with two comments on possible applications of the new
equations: A vector potential ${\bf A} ( {\bf r}, t)$ is included in
(\ref{t210}) by replacing ${\bf p}_{i} \rightarrow \bld{\pi}_{i}
= {\bf p}_{i}+ q_{i} {\bf A}({\bf r}_{i}, t )$. This allows one to calculate
relativistic recoil effects in positronium de--excitation, or in the
$e^{+}e^{-}$--recombination into positronium. In addition, the Lamb shift
may be calculated from these processes via a dispersion integral.

The second comment concerns the case $m_{1} = m_{2}\,$.
Here $M_{+}^{2} / s$ vanishes for the small components $\psi_{f}$
of $\psi\,$. 
Calling for brevity
\begin{equation}
{\widetilde V} = V ({\widetilde r} \,)\, ,\quad \widetilde{\bld{\pi}} =
{\widetilde {\bf p}} + \frac{i}{2} {\widetilde \nabla}\, {\widetilde V}
\bld{\sigma}_{1}\bld{\sigma}_{2} \, ,
\end{equation}
one has $\psi_{f}  = ( 1 - 2 {\widetilde V} ) ^{-1}\, 2 
\bld{\sigma}_{1}\widetilde{\bld{\pi}}\,  \psi_{g}\,$.
The large components satisfy the equation
\begin{equation}\label{letzt}
\left\lbrack 4 m_{1}^{2} / s - 1 + 2 {\widetilde V} +
4 \bld{\sigma}_{1} \widetilde{\bld{\pi}}\, (1-2 \widetilde{V} ) ^{-1}\,
\bld{\sigma}_{1} \widetilde{\bld{\pi}}
 \right\rbrack \psi_{g} = 0 \, . 
\end{equation}
For ${\widetilde V} = +\,  \alpha / {\widetilde r}\,$, it has only scattering
states, which would be $e^{-}e^{-}$ and $e^{+}e^{+}$ in
the case of two electrons.
If it were possible to extend the present formalism to an external
Coulomb potential, one would arrive at a theory which
isolates the $e^{-} e^{-}$-- and $ e^{+}e^{+}$--\,channels from the
$e^{-}e^{+}$\,--channel. This would dispense with {\sl positive--energy
projectors} and simplify the relativistic variational
calculation of atomic ground states.
\vspace{0.5cm}
\\
This work has been supported by the {\it Deutsche Forschungsgemeinschaft}.


\begin{references}
\bibitem{mapi}M. Malvetti and H. Pilkuhn, Phys.\,Reports {\bf 248}, 1
 (1994), and references therein
\bibitem{pi}H. Pilkuhn, J.\,Phys. B {\bf 28}, 4421 (1995)
\bibitem{wei}S. Weinberg, Phys.\,Rev. {\bf 133}, B1318 (1964)
\end{references}
\end{document}